\documentclass{article}
\usepackage{times}
\usepackage{RR}
\usepackage{hyperref}
\usepackage{a4}
\usepackage{graphicx}
\usepackage{amsmath}
\usepackage{epsfig}
\usepackage[]{latexsym,amssymb,amsmath, theorem}

\newcommand{\dd} { {\mathrm{d}} }
\newcommand{\bfrac}[2]{ { \displaystyle \frac{#1}{#2} } }
\RRdate{October 2008} 
\RRauthor{S. K. Godunov%
 \thanks{Sobolev Institute of Mathematics, Novosibirsk, Russia.}%
 \thanks{godunov@math.nsc.ru}%
 \thanks[sfn]{This preprint is a talk that was given at the
International Symposium entitled "Godunov's Method in Gas Dynamics",
Michigan University (USA), May 1997.}
}
\authorhead{S.~K.~Godunov} 
\RRetitle{Reminiscences about numerical schemes}
\RRtitle{Reminiscences about numerical schemes}
\titlehead{Reminiscences about numerical schemes}
\RRresume{Une version, \'ecrite en russe, est apparu la premi\`ere
fois en 1997. D'autres versions (anglaise et fran\c caise) on \'et\'e
publi\'ees mais sous forme abr\'eg\'ee. On pr\'esente ici une
traduction compl\`ete du document original en anglais, traduction
effectu\'ee par O. V. Feodoritova et V. Deledicque, envers lesquels
j'exprime ici toute ma gratitude.}

\RRabstract{This preprint appeared firstly in Russian in 1997. Some
truncated versions of this preprint were published in English and
French, here a fully translated version is presented. The translation
in English was done by O. V. Feodoritova and V. Deledicque to whom I
express my gratitude.} 
\RRmotcle{Sch\'ema de Godunov, syst\`emes hyperboliques de lois de
conservation.}
\RRkeyword{Godunov's Scheme, hyperbolic systems of conservation laws.}
\RRprojet{Smash}
\RRtheme{\THNum}
\URSophia
\begin{document} 
\RRNo{6666}
\makeRR
%
%
%
%
%
%
\section*{Introduction}

In the present paper I will describe how the first variant of the
"Godunov's scheme" has been elaborated in 1953-1954 and tell about
all modifications realized by myself (until 1969) and the group of
scientists from the Institute of Applied Mathematics in Moscow
(which has become the M.V.Keldysh Institute of Applied
Mathematics).

At the time these modifications (see Sections 2,3) were carried
out, other algorithms were developed, in particular  second order
schemes for gas dynamics problems with a small number of strong
and weak discontinuities [1-3]. We performed many calculations
based on the first codes written by V. V. Lucikovich. More
complicated problems resulted in an elaboration of  very artful
approaches to divide the whole computational domain into
sub-domains which have been developed by A.V.Zabrodin. This
procedure resulted in the ne\-ces\-sity to develop algorithms of
grid construction.

In 1961-1968  G.~P.~Prokopov and I carried out approaches to the
construction of moving grids which were used in serial
calculations by A.~V.~Zabrodin, G.~N.~No\-vo\-zhilova and
G.~B.~Alalikin (see [4-6]). The problems appearing in the grid
con\-struc\-tion forced us to solve elliptical systems (see
[7-8]). The methods  elaborated here, have later been employed in
elliptical spectral problems and have been presented in my papers
on numerical linear algebra (see [9,10]). The number of
interesting observations made during the analysis of my
calculations gave many discussions at the Mos\-cow University and,
after 1969~-- at the Novosibirsk University. As a result of such
dis\-cus\-sions the criterion of spectral dichotomy [11,12] has
been developed and high-precision algorithms to calculate singular
vectors have been constructed (see [11,13]). It is difficult to
imagine that the reason of such investigations has been the
elaboration of approaches to the gas dynamics calculus and
numerical grid constructions.

During my studies at Moscow University I learned  the differential
equations theory in seminars of I. M. Gelfand and I. G.
Petrovskii. The latter focused my attention on gas dynamics
problems and proposed to me to use  stationarization methods to
study transitional flows (with sub- and supersonic regions). My
qualification work was devoted to the stationarization  of a flow
inside a nozzle (however, only in subsonic regime and with
artificially added time derivatives introduced in Chaplygin's
equation). Petrovskii's idea about stationarization in practical
form was published in 1961 (see [6]). The technical statement was
presented in the qualification work of G. P. Prokopov performed
under my supervision. The coding was made by G. N. Novozhilova.

The elaboration of numerical schemes was carried out at the same
time with attempts to have a better understanding  of the notion
of generalized solutions to quasi-linear systems of equations. As
a rule, the hypothesis about possible definitions and properties
of the generalized solutions were preceded to the construction of
numerical schemes, which used these properties. At the same time,
I tried to prove the formulated hypothesis. To my deep
disappointment, these attempts had no success. On the contrary,
they often led to contradictory examples. But, at the same time, a
numerical scheme, more precisely its modification, based on using
the Euler coordinates, moving grids and tracking methods for
strong and weak discontinuous, was used in customary calculations.

\section{How the scheme has been elaborated}

In autumn 1953 M. V. Keldysh and I. M. Gelfand proposed me to
elaborate a variant of a method suggested by J. Von Neumann and R.
D. Richtmyer. It consisted in the introduction of artificial
viscosity in the gas dynamics equations. The goal was to finish
the algorithm for serial engineering calculations by the spring of
1954. By this time, the first electronic numerical device "Strela"
had to be delivered at our Institute.

I knew that one variant of the required algorithm  in our
Institute had been prepared by A. I. Zhukov.  I had the
opportunity to study it and the goal was to modify  this algorithm
or to suggest a new variant. From the point of view of the
administration of our Institute it was necessary to continue the
study in this direction  to be able to solve efficiently this
problem to a fixed date.

The method of A. I. Zhukov exactly coincided with the scheme
published by P. D. Lax one year later [14]. The paper I received
from A. I. Zhukov contained some ideas and hypotheses on which the
numerical scheme was based. Some of the hypotheses were erroneous
but during my first reading of the paper I was confident in them
and decided to rely on these hypotheses in my work.

The principal difference between Zhukov's method and the approach
by  J. Von Neumann and Richtmyer  lies in the use or not of the
artificial viscosity. A. I. Zhukov understood during his
experiments that smearing  of shock waves was made automatically
because of discrepancy between approximate and exact differential
gas dynamics equa\-tions. A. I. Zhukov explained this discrepancy
by analyzing the truncation terms of the difference scheme, which
present {\em numerical viscosity\/}, and formulated the idea
of {\em the first differential approximation\/}.

He began to use the second order difference schemes but faced with
strong oscillations near the shock wave front. To explain this
fact he used exact solutions based on Airy functions of linearized
equations in which numerical viscosity was introduced to model
truncation terms of the numerical scheme. The experiments resulted
in the conclusion that the second order schemes had to be avoided,
and A.~I.~Zhukov con\-cen\-trated on the development  of the first
schemes for the simplest quasi-linear Burgers equation and the gas
dynamics equations in the Lagrangian coordinates. The whole
attention was concentrated on the investigation of numerical
solutions of problems with a steadily  moving shock front. When
the first differential approximation was used, the problem was
reduced to the solution of the ordinary differential equations
which could be compared with numerical solutions. Based on these
comparisons A.~I.~Zhukov formulated a necessary condition which
any difference scheme should satisfy. This condition means that
the first differential approximation should have the form of
conservation laws, however it was formulated as a hypothesis of a
necessary and sufficient condition.

I did not doubt in that hypothesis and put it in the cornerstone
of my attempts to build a scheme without oscillations but with
second order accuracy. To reach this goal I began to compare
different second order schemes  which are not conservative but
they are conservative at the first order.  The comparison of
different variants was based on the same so\-lu\-tions which were
important for a whole set of typical problems. As a result a
rather satisfactory variant was chosen. Three months of intensive
work were spent for such a choice. N. M. Zueva helped to perform
calculus, and numerical experiments were performed by V. V.
Paleichik and two of her students who studied technology of
calculation based on mechanical adding machine "Mersedes". M. V.
Keldysh and I. M. Gelfand were interested in these investigations;
they listened to reports about the current state of this work
biweekly.

The next stage of the investigations was to prove the efficiency
of the scheme based on various numerical experiments  with
different spatial steps and different solutions.  But already
during investigation of the dependence of the solution on the
spatial step we faced with the situation that completely crossed
out all results of our three-month work.

To clarify the difficulties we have faced I have to tell about one
hypothesis from Zhukov's report. He compared the solution of a
problem about steadily moving shock waves by using the first
differential approximation with the real profile obtained from the
numerical calculations. They were different and he supposed that
this difference would be decreased if the spatial step tends to
zero.

I decided to check this hypothesis and asked V. V. Paleichik to
make cor\-res\-ponding calculations. One set of calculations with
one step size was made by the first student, and the calculations
with one-half of that step by another. Rapidly, in five-ten
minutes V. V. Paleichik reported me about senseless of the
proposed experiment. The results of both experiments were
absolutely the same, and obviously a different result was not
possible because spatial step itself did not play any role in
calculations. Only the relation between time and spatial steps
took part in the experiments and it was chosen constant to keep
Courant's number constant. So, finite-difference profiles were not
modified in the process of step decreasing and did not converge to
the dimensionless profile obtained for the first differential
approximation. At the same time, the comparison of the velocity of
the exact and finite-difference shock fronts demonstrated no
significant difference  $(\approx 3--4 \% )$, but this difference
did not decrease during steps reduction.

Obviously, this situation put me under stress and forced me to
change my previous point of view.

Firstly, it was obvious that the numerical scheme should guarantee
the correct ve\-lo\-city of a steadily moving shock wave and
correct values on both sides of the front {\em even if the
step size equals to 1}. The hope, that the accuracy in calculation
would be high if the step was small enough, disappeared.

The conclusion was to use exact  conservation laws without
simplifications related to the use of conservation laws obtained
from the first differential ap\-pro\-xi\-mation. For this, the
approximation of the equations
$$
\begin{array}{ll}
\displaystyle \frac{\partial u}{\partial t} + \frac{\partial p\,(v)}{\partial x}
& = 0\ , \\[.3cm]
\displaystyle \frac{\partial v}{\partial t} - \frac{\partial u}{\partial x}
& = 0\ ,
\end{array} 
$$
should be done to provide the following conservative form of
difference equations
$$
\begin{array}{ll}
\displaystyle 
\frac{u_n^{m+1}-u_n^m}{\Delta t}+\frac{P^{m}_{n+\frac{1}{2}} -
P^{m}_{n-\frac{1}{2}}}{\Delta x}& =0\ ,\\[.3cm]
\displaystyle 
\frac{v_n^{m+1}-v_n^m}{\Delta t} - \frac{U^{m}_{n+\frac{1}{2}} -
U^{m}_{n-\frac{1}{2}}}{\Delta x} & = 0\ .
\end{array} 
$$
To describe the velocity field one should use both $u_n^m$ and
$U^{m}_{n+\frac{1}{2} }$. Similarly, the specific volume $v$ and
pressure $p(v)$ related to it are approximated by both $v_n^m$ and
$P^m_{n+\frac{1}{2}}$ respectively. The different  formulas
connecting the lower and upper case letters result in the
different schemes. The resulting scheme should provide the
smoothness of a numerical profile. Based on this scheme and
numerical conservation laws one can conclude that velocity of the
wave is correct and values on the right and left sides of the
shock are connected by the known Hugoniot relations.

To construct a scheme I decided at first to obtain a smooth front
in the limit case of weak waves, i.e. in the case of acoustic
equations, which can be considered as the gas-dynamics equations
with the simplest equation of state 
$ p(v)=- a^2\, v\  (a^2=\hbox{\rm const.}>0)$.
In this case, the shock waves are transformed into the
discontinuities of the Riemann invariants $u \pm a\,v$, and the
equations are reduced to a canonical form
$$
\frac{\partial(u\pm a\,v)}{\partial t}
\mp 
a\,\frac{\partial(u\pm a\,v)}{\partial x}=0\ .
$$
I decided to choose numerical schemes which allow (at least in
this simplest case $p(v)=-a^2v$, $a=\hbox{\rm const.}$) the reduction
of these equations to independent equations for the Riemann invariants
$u\pm a\,v$. I used the simplest scheme conserving the monotonicity
of the Riemann invariants which was known and widely used at that
time. The monotonicity gua\-ran\-teed a required smoothness. At
that time I did not have another criterion and there was no more
time for proposing something else. "Strela" was already being
installed in our Institute. At that time I could show for the
linear case that only the first order schemes conserve the
monotonicity. I decided not to provide further efforts in order to
to construct second order schemes.

The chosen scheme in the simplest case had the following form :
$$
\begin{array}{ll}
\displaystyle 
\frac{u_n^{m+1}-u_n^m}{ \tau} 
- a^2\, \frac{v^m_{n+1} - v^m_{n-1}}{2\, h}
- a\, \frac{u^{m}_{n+1}
- 2\, u^{m}_{n} + u^{m}_{n-1}}{2\,h}& = 0\ ,
\\[.3cm]
\displaystyle 
\frac{v_n^{m+1}-v_n^m}{ \tau} - \frac{u^m_{n+1} - u^m_{n-1}}{2\, h} 
- a\, \frac{v^{m}_{n+1}- 2\, v^{m}_{n} + v^{m}_{n-1}}{2\,h}& = 0\ ,
\end{array} 
$$
and transformed into a  finite-difference conservation laws after
introduction of the following notations
$$
\begin{array}{lll}
P^{m}_{n+\frac{1}{2}} & = \displaystyle 
                         - a^2\, \frac{v^{m}_{n+1}+v^{m}_{n}}{2}
                         + a\, \frac{u^{m}_{n+1}-u^{m}_{n}}{2} 
                      & = \displaystyle 
                         \frac{p(v^{m}_{n+1})+p(v^{m}_{n})}{2} 
                         + a\, \frac{u^{m}_{n+1}-u^{m}_{n}}{2}
\ ,\\[.3cm]
U^{m}_{n+\frac{1}{2}} & = \displaystyle 
                          \frac{u^{m}_{n+1} + u^{m}_{n}}{2}
                         + a\, \frac{v^{m}_{n+1}-v^{m}_{n}}{2} 
                      & = \displaystyle 
                          \frac{u^{m}_{n+1}+u^{m}_{n-1}}{2}
                        - \frac{p(v^{m}_{n+1})-p(v^{m}_{n})}{2\,a}
\ .
\end{array} 
$$
It was decided to use the scheme in a general nonlinear case, but
the constant $a$ in formulas $P^{m}_{n+1/2}, U^{m}_{n+1/2}$  had
to be replaced by some, generally speaking, non-constant values
$a^{m}_{n+1/2}$. The first variant to calculate $a^{m}_{n+1/2}$
was the following
$$
\begin{array}{ll}
a^{m}_{n+1/2} & = \displaystyle \sqrt{-p'(v^{m}_{n}+v^{m}_{n+1})}
\ ,\\[.3cm]
a^{m}_{n+1}   & = \displaystyle \frac{1}{2}\,
                  \left( \sqrt{-p'(v^{m}_{n})}
                       + \sqrt{-p'(v^{m}_{n+1})}
                  \right)
\ ,\\[.3cm]
a^{m}_{n+1} & = \displaystyle \sqrt[4]{p'(v^m_n) p'(v^m_{n+1})}
\ .
\end{array} 
$$

These formulas were tested in the numerical experiments during
which the mo\-no\-to\-nicity of the shock profile was checked. The
initial data for the shock were chosen as constant values on the
right and left sides of the discontinuity. These constant values
were connected by Hugoniot relations.

During these experiments I began using other formulas for
$a^{m}_{n+1/2}$. In particular, I wanted to conserve  monotonicity
at the first step and I could construct some interpolation formula
for $a^{m}_{n+1/2}$ which gave the most acceptable results.

Here I should add some precisions to my story to correct the
simplifications I introduced. We soon began the above described
experiments  not with the simplest gas dynamics equations we used
before but with the system of three equations 
$$
\begin{array}{ll}
\displaystyle 
\frac{\partial u }{\partial t} + \frac{\partial p(v\, ,\, E)}{\partial x}
& = 0\ ,\\[.3cm]
\displaystyle 
\frac{\partial V }{\partial t} - \frac{\partial u}{\partial x} 
& = 0\ ,\\[.3cm]
\displaystyle 
\frac{\partial}{\partial t} ( E + \displaystyle \frac{u^2}{2} )
+ \frac{\partial(p\,u)}{\partial x} & = 0\ ,
\end{array} 
$$
with the following form of numerical conservation laws
$$
\begin{array}{ll}
\bfrac{u_n^{m+1}-u_n^m}{\Delta t}
+\bfrac{P^{m}_{n+\frac{1}{2}} - P^{m}_{n-\frac{1}{2}}}{\Delta x}
& = 0\ ,
\\[.3cm]
\bfrac{v_n^{m+1}-v_n^m}{\Delta t} - \bfrac{U^{m}_{n+\frac{1}{2}} -
U^{m}_{n-\frac{1}{2}}}{\Delta x} & = 0\ ,
\\[.3cm]
\bfrac{\left( E+\bfrac{u^2}{2} \right)_n^{m+1}
      -\left( E+\bfrac{u^2}{2} \right)_n^m}{\Delta t} 
      + \bfrac{ P^{m}_{n+\frac{1}{2}}\, U^{m}_{n+\frac{1}{2}} -
        P^{m}_{n-\frac{1}{2}}\, U^{m}_{n-\frac{1}{2}}}{\Delta x}
& = 0\ .
\end{array} 
$$

A selection of the interpolation formulas for $a_{n+1/2}^{m}$
became in this case a selection of variants to calculate
$P^{m}_{n+1/2}, U^{m}_{n+1/2}$ as a function of $u^{m}_{n}$,
$v^{m}_{n}$, $E^{m}_{n}$, $u^{m}_{n+1}$, $v^{m}_{n+1}$,
$E^{m}_{n+1}$. The adequate variant has been found after 2-3
weeks. It is important to remark that in the chosen formulas the
role of $a^{m}_{n+1/2}$ played the expressions
$$
\sqrt{\frac{(\gamma +1) P_{n+1/2}+(\gamma-1) p_{n}}{v_{n}}}
$$
which depended on the quantity $P_{n+1/2}$ (in these experiments
the equation of state $E=(p\, v)/(\gamma-1)$ was used). It
occurred to me that I saw such expressions somewhere in the
literature.

Not long before to that the book of L.~D.~Landau and
E.~M.~Lifshitz "The Mechanics of Continuous Media" was published.
In this book these quantities took part in formulas which
described the solution of the Riemann problem. I had to modify
slightly my interpolation formulas to obtain the final form of my
scheme. These modifications were significant  only in the case of
strong rarefaction waves. It happened in March 1954. At the end of
March, V. V. Lucikovich was included in our group to write a code
according to technical tasks that we composed by taking into
account useful recommendations of K. A. Semendjaev. At the end of
April I was on holidays for two weeks, and on the 5th May we began
our experimental and applied calculations.

On the 4th of November 1954 I defended my PhD thesis containing
the description of the suggested scheme. But this work was
published only in 1959. Just before, I tried to publish the work
in some journals without success. The journal "Applied Mathematics
and Mechanics" refused to publish it because it was purely
mathematical and without any relation to mechanics. In one
mathematical journal (I do not remember which one) the refusal was
motivated with the opposite statement. After that I. G. Petrovskii
as a member of the editorial group helped to publish it in the
journal "Mathematicheckii Sbornik" (see [15]).

I read the paper by P. D. Lax  [14]  only after my PhD thesis
defense. The scheme described in this work coincided with Zhukov's
scheme but without questionable hypotheses which were in Zhukov's
report and whose analysis helped me to con\-struct my scheme. If I
had received this paper one year earlier, the "Godunov's scheme"
would never have been constructed.

\section{Problems of approximation and effective accuracy}

Indeed, the work on modifications of the scheme invented in 1954,
was actively pursued. Consumers were very suspicious toward my
scheme.

In our Institute, more precisely, in  Steklov's Institute, from
which Keldysh's Institute was separated in 1953, the basic
hydrodynamic calculations were or\-ga\-nized by K.~A.~Semendjaev
much before I came. They were made with the help of the mechanical
adding machine "Mersedes" using the method of characteristics by a
large group of colleagues. The algorithm was thoroughly thought
over by K. A. Se\-men\-djaev and A. I. Zhukov. The basic attention
was payed to the table of results and the reduction of
intermediate records. This reduction was provided with memory
cells of the "Mersedes".  All results were presented in graphical
form and checked very carefully.

The technique of the calculations was absolutely automatic. My
female colleagues worked on machines as pianists and
simultaneously discussed their household and other typical
problems of female interest. They mocked especially young
researchers among whom I was. We were named by the contemptuous
word "that science". They supposed we built pseudo-scientific
theories, useless to overcome difficulties which only led to the
delay of calculations. One should remark, their salary depended on
the calculation volume which was defined by the number of table
rows without mistakes.

The graphical presentation of results on a graph paper had a very
high quality, and contained a visual plot of velocity, pressure
and density fields,  and also the trajectories of contact
discontinuities and characteristics giving the possibility to
follow the domain of influence of the initial data.

Our customers - physicists and engineers~-- had a habit to a
perfect form of the information obtained and it was a reason of
their displeasure toward the first calculation results based on
numerical schemes. Now it is well-known that the numerical
calculations of the discontinuous gas dynamics solutions present a
"numerical" micro-structure which looks like errors providing a
caricatural picture of the real flow. The comparison of computer
and "hand-made" calculations resulted in the opinion that the code
had errors or that the initial data were erroneous. We had to
check carefully all situations, we had to explain the results and
to modify either the results or the calculation scheme.

I will describe now two such examples, the most interesting from
my point of view.

During calculations based on my scheme of strong isentropic
rarefaction waves  one could observe a significant growth of the
entropy which was the source of distrust toward the numerical
results. One noticed a substantial reduction of such effect only
after 3-4 years (in 1957-1958) when we studied 2-D problems. Such
problems, as a rule, do not have a simple formulation in the
Lagrangian coordinates and we had to use the Eulerian description.
Simultaneously, numerical grids became complicated, they became
moving ones. The main idea - to use the exact solution of the
Riemann problem as an element of a numerical scheme - has been
applicable in 2-D problems also, but not in a transparent way. The
first test calculations of 1-D problems earlier computed were
repeated using new 2-D codes. We were very surprised by the fact
that 2-D codes based on the Eulerian coordinates resulted in
essentially less parasitic increasing of the entropy in the
rarefaction waves.

The reason of the effect is that the Lagrangian coordinates (even
in the simplest 1-D case) are unfit to the formulation of the
generalized solutions of the gas dynamics equations. These
equations admit the formation of vacuum regions which are treated
in the Lagrangian coordinates by singularities of the  delta
function type. Usually such singularities are not supposed to be
present, and they are modelled very badly numerically. The
analysis of the results showed that even a sufficient decreasing
of the step size in Lagrangian coordinate in a strong rarefaction
zone does not result in a noticeable distance reduction between
Eulerian coordinates of neighboring nodes. I used this fact in
1961 in the theoretical work [16] to prove that my scheme
approximates (if the Eulerian coordinates are used) the gas
dynamics equations  in a sense of conservation laws (1-D case). At
the beginning of this work I hoped to prove a convergence to the
exact solution but I failed to that. Even the order of
approximation appeared to be different from one, it was  equal to
2/3. Remember that, formally, my scheme has the first order of
approximation based on the first differential approximation. I
should remark that indeed the order of approximation is probably
higher than 2/3. Apparently, based on estimates of the  variations
of solution suggested by J. Glimm [17] one can prove that this
order is equal to one. Unfortunately I did not investigate this
issue in details. However at the end of the 50th I was interested
basically in the accuracy of the approximated solutions, and not
in the problem of approximation of equations (i.e. what power  of
the spatial step evaluates the error - the  difference between the
calculation result and the exact solution). The obtained estimates
for approximation played a preliminary role for theoretical
investigation which I could not finish. Despite of it, in
1957-1958  V.S.Ryabenkii and I carried out an experimental study
of the convergence of approximate solutions calculated with my
scheme to the exact solution. These observations about calculation
results allowed me to conclude that the convergence had to be
considered as {\em weak\/} and not {\em strong\/} to exclude the
influence of significant deviations in one-three nodes. As a rule, the
occurring of these deviations is related to the interaction of two
shock waves or reflection of a shock wave from a contact
discontinuity.

During our discussions, V. S. Ryabenkii suggested an approach
which permitted us a reduction of the study of weak convergence to
the study of strong convergence {\em  in~\hbox{1-D}\- 
pro\-blems\/}. For this, one needs to compare not quantities controlled by
conservation laws but integrals (or even multiply integral) of
these quantities with respect of spatial coordinate. We supposed
to demonstrate the first order of weak convergence, i.e. the order
corresponding to a formal order of approximation. We were very
surprised to see that the observed effective convergence order was
less than 1 in experiments with centered expansion  waves or large
gradients  in smooth domains restricted by strong discontinuities.

At that time we were very busy and had no motivation to prepare a
detailed paper, especially because the conclusions were unpleasant
to us (to me at least) and did not correspond to our expectations.
I only made a brief presentation at a scientific conference at the
Moscow University. From my point of view, nobody put attention on
this presentation except N. S. Bachvalov who began the theoretical
study of the convergence rate for the solutions for the Burgers
equation  as $\varepsilon \rightarrow 0$
$$
\bfrac{\partial u }{\partial t} 
+ u\, \bfrac{\partial u}{\partial x}
= \varepsilon\, \bfrac{\partial^2 u}{\partial x^2}\ ,
$$
and obtained the estimate of the error $\varepsilon
 \ln |\varepsilon |$. At the same time,  all my attempts to attract
attention of researchers and engineers to these effects had no
success.

My colleagues A. V. Zabrodin, K. A. Bagrinovskii, G. B. Alalikin
and I were in\-volved at that time in preparing 2-D calculations.
That is the reason why the question  if the first order scheme has
the first order convergence rate, or, more precisely, what is this
order, was not understood.

I know that later  B. van Leer, P. Woodword, P. Collela, P.D. Lax,
A. Harten and many others (see for instance [18-21]) suggested to
construct efficient numerical schemes with order of approximation
higher than one. At that time I had already left Keldysh Institute
in Moscow and moved to Novosibirsk. Here I was involved in other
problems. Among these issues the basic place was devoted to
problems which arose during my work in Moscow and they required
deep theoretical in\-ves\-ti\-gations. They were, in particular,
issues concerning an accurate mathematical formulation of the
conversation laws and thermodynamical identities, energy
in\-teg\-rals for the hyperbolic equations, the mathematically
correct formulation of the elasticity theory, and the statement of
problems in numerical methods of linear algebra.  I had already
mentioned that this latter issue appeared from attempts to
construct 2-D numerical schemes. The investigation of these
various and interesting problems (see for instance [22,23]) did
not allow me to continue the study of numerical methods and a deep
understanding of  modern numerical schemes for the gas dynamics
equations, some of which named {\em "second order Godunov's schemes"\/}. I
claim I cannot be considered as the author of them. Maybe, their
authors were inspired by my suggestion about using the exact
solutions of elementary problems to construct a numerical scheme.
It is my pleasure to thank them for the attention they payed to
me.

Recently, two years ago, I learned that V. V. Ostapenko
(Lavrentiev Institute of Hydrodynamics SB RAS, Novosibirsk)
studied actively the problem of an effective accuracy of numerical
schemes in hydrodynamics based on asymptotic expansions. Instead
of performing cumbersome ana\-ly\-tical calculations, I advised
him to make the experimental investigations of the scheme we
constructed together with V. S. Ryabenkii in 1957-1958. I hoped he
would repeat our experiments and they would be described at last
in details. But V. V. Ostapenko concentrated on the investigation
of a more modern Lax-Harten scheme [19,21]. This scheme also
demonstrated a mismatch between the formal order of approximation
and the effective order of accuracy (order of convergence). At the
same time, from the experimental results of V. V. Os\-ta\-pen\-ko
it apparently followed that the accuracy of the Lax-Harten scheme
in the sense of weak convergence on discontinuous solutions is not
less than the first order, i.e. the accuracy is higher than for my
original scheme.

I think, it is very interesting to carry out massive experimental
investigations of precision for  all principal numerical methods.
Especially interesting would be such a study for 2-D and 3-D
problems which require the elaboration of  the corresponding
experimental techniques.

One should remark the active work of K. A. Semendjaev to transform
my scheme into a final computer code. His experience was very
useful for our success.

\section{2-D problems and moving grids}

Despite the fact that the work on the 1-D approach has been
continued I started thinking on a 2-D variant of my algorithm. The
attempts to solve gas dynamics problems were done in Steklov's
Institute and later in our Institute by K. I.~Babenko and I.
M.~Gelfand. Appearing difficulties were discussed on seminars
where M. V.~Kel\-dysh took part. In particular during these works
and accompanying discussions the numerical schemes~-- explicit in
one spatial variable and implicit in the other one ("sausages")~--
appeared. Also the first variant of the matrix double-sweep method
that was suggested by M. V.~Keldysh and investigated by K.
I.~Babenko and N. N.~Chentsov appeared. There were a lot of
discussions after the Lo\-cut\-sievskii's report based on the
Ray\-leigh's book "Theory of Sound" chapter devoted to the
instability of a contact discontinuity. Indeed, the contact
discontinuity evolves in time according to the law described by
differential equations for which the Cauchy problem is not well
posed in Hadamard's sense ( and not only its solutions  are
instable). The non well-posedness  is characterized  by the
following fact~-- the short waves increase their amplitude very
fast, the smaller the wave length is, the faster it is. The
instability is a similar phenomenon, but in this case the
characteristic time, during which the amplitude of the wave is
in\-creased, is bounded for all wave lengths. It was thought that
processes described by the equations which are non well-posed in
Hadamard's sense could not be computed numerically because
numerical schemes should be unstable. We discussed some
regularizations of the boundary conditions on the contact
discontinuity. However I cannot remember if any realistic variant
of such a regularization was developed.

Impressed by these discussions I got into a panic about the
problems where contact discontinuities appeared. At that time
there was a large interest in the problems that could be
considered as near 1-D, i.e. the problems with slightly curved
shocks and sound fronts and almost plane contact discontinuities.
It occurred to me that for these problems a simple generalization
of my scheme could be applied, and I started developing such
generalization. My solution was supported by M. V.~Keldysh and I.
M.~Gelfand. K. A.~Bagrinovskii and G. B.~Alalykin took part in the
development of the first variant of algorithm. Later, A. V.
Zabrodin joined us. The codes were again developed by V.
V.~Lucikovich and G. N.~Novozhilova. It is interesting to note
that, when the 2-D computer code has been written, the problems
became more complex due to increasing demands of engineering
interests. These complications led to modifications of the code
and of the elements of the computational method. Busy with the
computer code we forgot about the main danger~-- non
well-posedness of the contact discontinuity. As far as I remember
this non well-posedness did never appear.

I cannot understand why we did not meet this non well-posedness.
Apparently, finite-difference equations used for the calculation
of the boundary trajectories provide some forced regularization.
If it is really so, one should be anxious about the following. The
mechanism of this regularization is unknown, an hence we are not
able to guarantee that it does not produce some undesirable
effects.

From my point of view, {\em the detailed analysis of the
algorithms used for the calculation of contact
discontinuities is still an actual problem\/}. In this
case, we can expect that some new effects will be discovered.

One should note that calculations with moving grids, in which the
tracked shock wave induces the displacement of points in 2-D and
3-D grids, has not been studied carefully, neither theoretically
nor experimentally. The feedback of such point displacements has
not been investigated. And it is not clear what kind of effects
this feedback produces.

When we implemented the 2-D approach based on the solutions of the
Riemann problem with arbitrary initial conditions, the first
question concerned  the construction of such solutions. In the 2-D
case the rectangular grid cells can neighbor not only on each
other but also on the nodes where four cells meet. If one
constructs a 2-D scheme analogously to  the  1-D, one should have
analytic solutions of hydrodynamical equations with four
discontinuities of initial data at one point. We did not have such
solutions, even now they do not exist, at least for general
initial data. We had the audacity to  suggest to use only
classical solutions of the Riemann problem combined from the plane
waves and describing initial Riemann problem placed on the edges
of the neighboring cells. We ignored interaction of four cells
having a common node. Implementing this approach we abandoned a
clear physical interpretation on which the construction of the 1-D
scheme was based. Obviosly, there were a lot of discussions about
the suggestion mentioned above. As a result we decided to perform
the investigation with the help of a calculation of a shock or
even an acoustic wave moving in the direction of a grid cell
diagonal. At the same time, for the acoustic waves propagating in
a medium  at rest, K. V. Brushlinskii, based on Gelfand's
suggestion, constructed the solution of the problem using an
interaction of all cells adjoint to one node with the help of the
Sobolev's method of functionally invariant solutions. This
solution was used in a numerical scheme completely analogous to
the 1-D one. After that, the calculations of waves moving in a
grid diagonal direction were performed with the help of
Brushlinskii's scheme and compared with our scheme. To our
surprise and satisfaction, we did not discover any essential
differences. After that, only the rough model was employed. We
made a lot of efforts to derive a stability criterion needed to
choose an admissible time step for a prescribed spatial step.
First, using the Fourier method,  G.B.~Alalykin, K.A.~Bagrinovskii
and I, arrived at a cubic characteristic equation and derived from
it only the necessary  condition of stability. Then, together with
K.A.~Bagrinovskii, we succeeded in obtaining a sufficient
condition of stability considering results of 2-D calculation as
some averaging of 1-D calculations, or as one says by using a
{\em splitting\/} approach. This work [24] was published in 1957 where a
few not very simple tests were performed with the help of our 2-D
code.

The idea to use the {\em splitting\/} approach appeared under the
influence of investigations of 2-D implicit schemes for  the heat
equation. These schemes were analogous to ones that were proposed
later by J. Douglas and their co-authors. But at that time we
refused to use {\em splitting\/} for the 2-D heat equation.

Similarly to the construction of the  numerical schemes for the
gas dynamics and acoustics using non-smooth~-- discontinuous~--
solutions, I decided to test proposed splitting schemes for the
equation $u_t=u_{xx}+u_{yy}$, taken as a test for the problem of
the evolution of a heat impulse released in one grid cell. In
other words, I decided to simulate the solution which  after some
time should not be essentially different from
$$
u(x,y,t)=\bfrac{\hbox{const}}{t}\,\exp\left(-\bfrac{x^2+y^2}{4t}\right)
\ .
$$
At least, the level curves of the computed solution should be
convex curves~-- close to circles.

I was very surprised when, during calculations with the Courant
number
$$
\bfrac{\Delta t}{(\Delta x^2+\Delta y^2)}\approx 10\ ,
$$
I realized that these level curves turned out to be cross-like at
least at the first time steps. This phenomenon forced me to avoid
the splitting for the heat equation. However, these investigations
led us to use the splitting procedure not only for the stability
analysis of gas dynamics numerical schemes but also for the
organization of the code structure.

The first 2-D code was constructed in such a way that on the
successive steps the fluxes, computed with the help of the 1-D
approach, were used at first in one direction and then in the
other grid direction. There were no troubles, but later we refused
to use splitting in these problems with explicit schemes as V. V.
Lucikovich suggested. He said that this refusal allowed to
simplify the code and made it more fast-acting. Before we were
sure that using the splitting approach led to simplification.

As I mentioned above we had to use the Lagrangian coordinates and
turned to the Eulerian coordinates. But at the same time in order
to connect the grid with moving boundaries, we had to make the
moving grids as well. In particular, it allowed to include the
shock fronts into a special type of boundary. The usefulness of
this fact was demonstrated at the calculation of the flow around a
sphere [6].

One should remark that in the first variants of our 2-D approaches
we constructed meshes and wrote finite-difference formulas on the
basis of conservation laws with a lot of caution and carefulness.
For example, for cell boundaries, the arcs of logarithmic spirals
were used,  and the integrals over the cells limited by these
spirals were calculated by explicit formulas. Only some years
later we began using more simple variants. The hard work to derive
the analytic formulas for integrals was performed by K.
A.~Bagrinovskii and A. V.~Zabrodin.

Even in 1-D calculations that were performed in the Lagrangian
coordinates we tried to increase the  accuracy by looking at the
advancement of the strongest shock waves. In order to avoid the
spreading of these waves,  their coordinates were marked and the
grid cells where they were located were divided into two parts. To
compute the displacement of a particular wave, the Hugoniot
relations were used. After that, we started using the moving grids
in the Eulerian coordinates, and the necessity to distinguish the
boundary type~-- shock waves and the boundaries of material layers
(contact discontinuity)-- disappeared. It simplified  the logical
structure of the algorithm and allowed us to introduce one more
boundary type on Zabrodin's suggestion. These boundaries had an
assigned position  or moved following a specified law and did not
influence the medium flowing across them. They were called
Eulerian boundaries.

The grids in regions, limited by boundaries of different types,
were attached to these boundaries and computed with the help of
simple interpolation formulas. Of cause, at each time step, the
displacement of boundaries  implied the displacement of grid
points. The use of codes with such a multi-region structure
allowed to increase the accuracy. It was possible not due to the
fact that numerical formulas were improved, but because the grid
adapted to the solution's structure and strong discontinuities
were not smeared. Working with this code I wanted to change the
grid calculation based on the interpolating formulas in a
geometric region with assigned boundaries into the solution of
some differential equations. These equations described a mapping
transforming this region into some standard one (for example, into
rectangular).  G. P.~Prokopov and I have studied this problem for
seven years (1961-1968) to achieve the first acceptable results
[5,25,26].

I am still interested in the classes of mappings for the grid
generation although I stopped studying numerical schemes many
years ago [27].

From the beginning,  G. P. Prokopov and I postulated that the
mapping should be defined as a solution of elliptic equation
systems. Our main efforts were directed to solve them efficiently.
We decided to use a variational approach that was realized with
the help of a finite element method. However, we did not perform
the first test on calculation of grids in hydrodynamical problems
with moving boundaries. Because in this case the elliptic system
should be solved on each time step (we were afraid of expensive
time-cost). First, we decided to use a variational approach in
some stationary problems and applied our ideas to calculate the
critical parameters of a nuclear reactor (see [8]). I mentioned
above that during these secondary investigations I was interested
in computational methods of linear algebra. I was attracted by
this problem, and I devoted to it my whole attention.

Only after that we succeeded in finding the solution of stationary
problems, we decided to generate grids on  each step  by solving
elliptic equations. The first sufficiently universal code started
to work at the end of 1968 -- at the beginning of 1969. The first
problem computed by it was based on the article [28]. This article
contained the talk presented at the International conference on
explosion physics (Novosibirsk, 1969)  and was devoted to the wave
formation at the explosion welding and the analysis  of
experiments that were carried out at the Novosibirsk Institute of
Hydrodynamics.

In September 1969, I moved to Novosibirsk and later did not devote
myself to numerical schemes.

I should remark that during the development of the moving grids
for 2-D calculations, we tried to use them in 1-D hydrodynamical
problems to obtain more exact results. The whole domain  was
divided into some sub-domains whose boundaries were contact and
shock discontinuities, and characteristics. Inside each sub-domain
a scheme of the second order was used. It was important for the
calculation of rarefaction waves to use  many nodes immediately
after they emerged (for instance, 50). It provided very high
accuracy, and it was possible because of application of implicit
numerical schemes.

We carried out the described method together with I. L. Kireeva
and L. A. Pliner, and the algorithm and code were written by G. B.
Alalikin. Based on this code some principal calculations were
made. The method was published  in 1970 (see book [1], in
russian). I think the book passed unnoticed but I believe the
specialists in the area of numerical hydrodynamics can find
something interesting even today.

One should mention that at the same time I worked with the
characteristics method to adapt it to computer calculations. The
result of this activity was the method of characteristics by
layers in which not the intersection points of characteristics are
calculated, but the coordinates  of these characteristics at fixed
layers $t=\hbox{const}$.  But after having finished both the method and
the code, during the first calculations the method showed its
inadequateness.  It did not calculate strong rarefaction waves
correctly. We did not expect such a situation because the
characteristics method  supposed to be good in calculation of such
waves. Brief communication about this failure was published in [3]
and it led me to construct a scheme of the second order to
calculate solutions with distinguished discontinuities to which
the above cited book is devoted [1].

The scheme for the gas dynamics calculations based on the solution
of the Riemann problem  began wide-spreading only in 1969, where
even three independent presentations were reported during the
conference in Novosibirsk. I have already mentioned our
presentation. Besides, there was the presentation of M. Ya. Ivanov
and A. N. Kraiko from Moscow Central Institute of Aircraft Engines
and the pre\-sen\-tation of T. D. Taylor and V. S. Mason (USA)
about using our scheme to calculate gas-dynamical flow around
"Apollo" (see [29,30]). Later, together with our colleagues from
Institute of Aircraft Engines we prepared a detailed description
of the numerical method and published the book translated in
French [31].

\section{Conservation laws and thermodynamics}

I will touch one more issue appeared during the work on the
construction  of numerical schemes and related to the concept of
generalized solutions of conservative qua\-si\-linear equations.

As it was mentioned above the conservation laws of mass, momentum
and energy are valid for the numerical solutions computed with the
help of my scheme. However the classical solutions (without
discontinuities) of gas dynamics equations obey one more
additional conservation law~-- the entropy conservation law. The
entropy increases in the case where gas passes through the shock
front (dis\-con\-tinuity in the solution). This statement is the
Zemplen's theorem, which represents (from the point of view of the
theory of quasi-linear equations) the postulate included into the
definition of generalized solutions. With the help of this
postulate the discontinuous solutions satisfying the conservation
laws of mass, momentum and energy, so that the entropy of a given
material volume is decreasing, are excluded from the number of the
generalized solutions. In my scheme, the law of non-decreasing of
the entropy is automatically fulfilled because the solutions of
the Riemann problem included in this scheme satisfy this law, and
also  the entropy is increasing in the process of averaging in
computational cells. Such averages are performed at the end of
each time step in my scheme.

Obviously, I was interested in the description of such
quasi-linear equations for $n$ unknowns whose classical solutions
satisfy automatically to a $n+1$-th conservation law. Trying to
understand  which is the answer on this question, I noticed that
all nonlinear relations which are present in the  hydrodynamical
equations are defined by only one nonlinear func\-tion~-- the
thermodynamic potential $E(V,S)$. In the Gelfand's lectures on the
theory of quasi-linear equations that he gave at the Moscow
University [32], the role of in the \-equa\-li\-ties $E_S>0$ and
$E_{VV}\,E_{SS}-E^2_{VS}>0$ was mentioned. These inequalities in the
thermodynamics are used for the gases and liquids if the
parameters $V$, $S$ are far from the phase-transition values. As a
problem,  I. M.~Gelfand suggested to prove that these inequalities
provide the well posedness  of spreading of sound-waves by
acoustic equations for a heat conducting gas. Impressed by this
problem I supposed that similar considerations about well
posedness  or stability (for systems with finite numbers of
degrees of freedom) can be put in the base of a well-known
thermodynamic theorem on the existence of the integrating
multiplier $1/T$ for $\dd E+ p\, \dd V$. This theorem is the  basis of
the entropy definition with the help of the equality
$\dd S= (1/T)\,(\dd E + p\, \dd V)$. After several months of hard work
I succeeded to prove this hypothesis [33]. I found out that the
stability, that proves the impossibility to construct  the perpetual
motion machine of the second kind, is always the consequence of
existence of some integral which is dissipated due to the heat
transfer. If we study only the stability of small oscillations
(i.e. describing the evolution of a system by linear equations), the
dissipating integral appears to be a quadratic form with a symmetric
and positive definite coefficient matrix. The symmetry of this matrix 
leads to the equalities from which the existence of a universal
integrating multiplier is derived. In my article [33] (see also
\S~22 in the book [23]) the drafts of the perpetual motion machine
of the second kind are presented in the case where the universal
integrating multiplier does not exist.

After finishing the work [33], in order to understand the reasons
for which the three equations have the "extra" fourth conservation
law, it was natural to represent the hydrodynamical equations in
the  Lagrangian coordinates (they are used for classical
solutions)
$$
\begin{array}{c}
\bfrac{\partial V}{\partial t} + \bfrac{\partial u}{\partial x} 
= 0\ ,\\[.3cm]
\bfrac{\partial u}{\partial t} + \bfrac{\partial E_V(V,S)}{\partial x}
= 0\ ,\\[.3cm]
\bfrac{\partial S}{\partial t} = 0\ ,
\end{array} 
$$
and derive the fourth conservation law as their linear combination
$$
\begin{array}{c}
0 = \bfrac{\partial(E(V,S) + u^2/2)}{\partial t}
+ \bfrac{\partial( u\, E_V(V,S))}{\partial x}
\\[.3cm]
\equiv E_V(V,S)\, 
\left[ \bfrac{\partial V}{\partial t} 
+ \bfrac{\partial u}{\partial x} \right]
+ u\, \left[ \bfrac{\partial u}{\partial t}
+ \bfrac{\partial E_V(V,S)}{\partial x} \right] 
+ E_S\, \bfrac{\partial S}{\partial t}\ .
\end{array}
$$
Obviously, such an equality representing a linear dependence
between four conservation laws can be solved relatively to any of
them. In particularly, the entropy conservation law
$\bfrac{\partial S}{\partial t} = 0$ can be presented as a
linear combination of the conservation law for the specific volume
$$
\bfrac{\partial V}{\partial t}
+\bfrac{\partial u}{\partial x} = 0\ ,
$$
the conservation law of the momentum
$$
\bfrac{\partial u}{\partial t}
+ \bfrac{\partial E_V}{\partial x} = 0\ ,
$$
and the conservation law of the energy in the following way
$$
\begin{array}{c}
0 = \bfrac{\partial S}{\partial t} =
- \bfrac{E_V}{E_S}\, \left[
 \bfrac{\partial V}{\partial t}
+\bfrac{\partial u}{\partial x} \right]
- \bfrac{u}{E_S}\, 
\left[ \bfrac{\partial u}{\partial t} 
+ \bfrac{\partial E_V}{\partial x} \right] 
+ \bfrac{1}{E_S}\, \left[
 \bfrac{\partial ( E + u^2/2 )}{\partial t}
+\bfrac{\partial ( u\,E_V )}{\partial x}
                  \right]\ .
\end{array}
$$
The last equality is possible due to the relation between the
differentials
$$
\dd S = - \bfrac{E_V}{E_S}\, \dd V 
        - \bfrac{u}{E_S}\, \dd u
        + \bfrac{1}{E_S}\, \dd( E + \bfrac{u^2}{2})\ .
$$
It is convenient to use the following form
$$
\dd \left[ S + \bfrac{E_V}{E_S}\, V
             + \bfrac{u^2}{2\,E_S}
             - \bfrac{1}{E_S}\,( E + \bfrac{u^2}{2} )
   \right]
= V\, \dd \bfrac{E_V}{E_S}
 + u\, \dd \bfrac{u}{E_S}
 - ( E + \bfrac{u^2}{2} )\, \dd \bfrac{1}{E_S}\ ,
$$
and to introduce the notation
$$
\begin{array}{ll}
L & = S - \bfrac{1}{E_S}\, ( E + \bfrac{u^2}{2} )
      + \bfrac{V\, E_V}{E_S}
      + \bfrac{u^2}{2\,E_S}
  = \bfrac{1}{E_S}\, [ S\, E_S + V\, E_V - E ]\ ,
\\[.3cm]
q_1 & = \bfrac{E_V}{E_S}\ ,\ \quad 
q_2 = \bfrac{u}{E_S}\ ,\ \quad
q_3 = - \bfrac{1}{E_S} = - \bfrac{1}{T}\ .
\end{array} 
$$
As a result, we have the equalities that relate derivatives
$L_{q_i}$ to the quantities appearing in the hydrodynamical
equations under the derivative $\bfrac{\partial}{\partial t}$
$$
L_{q_1} = V\ ,\ \quad L_{q_2} = u\ ,\ \quad 
L_{q_3} = - ( E + u^2/2 )\ ,
$$
and the relations between the differentials $\dd S$, $\dd V$, 
$\dd u$, $\dd ( E + u^2/2 )$ in an elegant form, on which the
well-known Legendre's transform in the theory of convex functions is
based
$$
\dd( q_1\, L_{q_1} + q_2\, L_{q_2} + q_3\, L_{q_3} - L )
= L_{q_1}\, \dd q_1 + L_{q_2}\, \dd q_2 + L_{q_3}\, \dd q_3\ . 
$$
It is convenient to introduce one more function of $q_1$, $q_2$,
$q_3$
$$
M = - \bfrac{q_1\,q_2}{q_3}\ ,
$$
such that
$$
\begin{array}{l}
M_{q_1}  = - \bfrac{q_2}{q_3} = u\ ,
\\[.3cm]
M_{q_2}  = - \bfrac{q_1}{q_3} = E_V\ ,
\\[.3cm]
M_{q_3}  = - \bfrac{q_1\,q_2}{q^2_3} = u\, E_V\ ,
\\[.3cm]
M - q_1\, M_{q_1} - q_2\, M_{q_2} - q_3\, M_{q_3} = 0\ .
\end{array} 
$$
In this case the hydrodynamical equations have the
following form
$$
\bfrac{\partial L_{q_i}}{\partial t} 
+ \bfrac{\partial M_{q_i}}{\partial x} = 0\ ,\ \qquad i=1,2,3\ .
$$
They consist of three conservation laws, and the forth (additional)
one
$$
\begin{array}{c}
\bfrac{\partial}{\partial t} 
      ( q_1\, L_{q_1} + q_2\, L_{q_2} + q_3\, L_{q_3} - L )
- 
\bfrac{\partial}{\partial x}
      ( q_1\, M_{q_1} + q_2\, M_{q_2} + q_3\, M_{q_3} - M ) 
\\[.3cm]
\equiv 
\bfrac{\partial}{\partial t}
       ( q_1\, L_{q_1} + q_2\, L_{q_2} + q_3\, L_{q_3} - L)
= 0\ ,
\end{array}
$$
coincides with the conservation law of entropy $\bfrac{\partial
S}{\partial t} = 0 $. Afterward it was not
difficult to obtain for other classical equations of mathematical
physics, even multidimensional, the standard representation
$$
\bfrac{\partial L_{q_i}}{\partial t} + \bfrac{\partial L^j_{q_i}}
                                             {\partial x_j} = 0\ .
$$
I succeeded in writing in a similar form the gas dynamics
equations in the Eulerian coordinates. I was happy with the fact
that if the generating {\em "thermodynamic potential"\/} 
$ L = L(q_1,q_2,\ldots)$ is convex, these equations can be rewritten
in the form of the Friedrichs's sym\-met\-ric hyperbolic system
$$
L_{q_i q_k}\, \bfrac{\partial q_k}{\partial t}
+ L^j_{q_i\, q_k}\, \bfrac{\partial q_k}{\partial x_j} = 0\ ,
$$
with a positive definite matrix of coefficients at derivatives
$\partial q_i$ on $t$. K. O. Friedrichs showed that for such
systems the Cauchy problem is well-posed  if some con\-di\-tions
on the smoothness of initial data are true.

So, my hopes to discover the connection between well-posedness and
the laws of phenomenological thermodynamics have been realized.

I described all my achievements in the article that was sent to
the journal "Uspekhi Matematicheskikh Nauk", but it was not
published because a reviewer decided that it did not have a
mathematical content.

After that I was dealt a new severe blow. I decided to use the
obtained equations by introducing into them small additional
dissipation terms. I wanted to confirm that the discontinuous
solutions, after introducing any small viscosities, are just
"smeared" in a thin stripe, and that the limit solutions
(generalized solutions of equations with zero viscosity) do not
depend on the form of these viscosities. Considering solutions in
the form of travelling waves $ q_i = q_i(\xi)$, $\xi = ( x -\omega\,
t)/\varepsilon$ for equations 
$$
\bfrac{\partial L_{q_i}}{\partial t} 
+ \bfrac{\partial M_{q_i}}{\partial x} 
= \bfrac{\partial}{\partial x}\, 
 \left[ \varepsilon\, b_{ik}(q)\, \frac{\partial q_k}{\partial x}
\right]\ .
$$
I  could propose an obvious geometric interpretation for ordinary
differential equations for  $q_i(\xi)$. This interpretation led to
a construction of systems of hyper\-bolic conservation laws for
which admissible discontinuous solutions satisfied the entropy
increasing law,  and which depend on what kind of small
dissipative terms we introduced to smooth them. In physical
problems the real dissipative processes can differ from the
numerical dissipation that depends on the numerical scheme. Soon,
B. F.~Diachenko constructed the example in which different
numerical approximations led to different numerical solutions
[34].

I was very upset that I could not justify all hypotheses lying in
the background of the numerical scheme. My work with the
description of this example was  presented in "Doklady AN SSSR"
[35], [36] by I. G. Petrovskii. Before the article was
pub\-lished, I presented it on the seminar  where R. Courant and
P. D. Lax participated, and visited Moscow for the first time.

Later, in Novosibirsk, together with my colleagues I continued the
investigations on conservation laws and their connections with the
thermodynamics. The review of a part of these investigations was
included into the paper [22] that I presented in 1986 in
St.-Etienne at the Conference on hyperbolic equations. I should
also mention important studies [37], [38] on this issue performed
by P. D.~Lax and K. O.~Friedrichs. Last years these studies were
continued and presented by me and E.I.Romenskii in Lisbon, Lake
Tahoe (USA, Nevada), Paris [39-41]. Our work [42] (in
collaboration with T. Yu. Mikhailova) was also devoted to this
topic. The works [41],[42] showed unexpected connection between
formally overdetermined systems of conservation laws in
mathematical physics and the theory of representations of the
rotation group.

I should notice a recent work [43] in which the estimates of the
entropy growth for the systems of  conservation laws in the form
described above are investigated. The author of this work could
connect these estimates with the {\em variation\/} of solutions. It
seems to me that this original way of definition of such a
{\em variation\/} is very promising. For this definition it is not
important if the problem is one-dimensional or multi-dimensional.

\section{Conclusion}

In this note I tried to describe the intensive work in the field
of numerical hydrodynamics in which I was involved by my
supervisers I. G. Petrovskii and I.~M.\- Gel\-fand, and the result was
the development of  {\em Godunov's scheme\/}.

I should emphasize the influence of the whole scientific community
within I worked. Of course, different approaches to the numerical
hydrodynamics were elaborated  by  many other research groups and
some of these groups belonged to the Institute where I worked. The
scientists who participated in these groups are K.~I.~Babenko,
V.~V.~Rusanov, I.~M.Gel\-fand, V.~F.~Diachenko, A.~A.~Samarskii,
N.~N.~Janenko, B.~L.~Rozh\-des\-tvenskii, and others. My goal was
not to describe the whole history of this subject but only present
the part in which I participated.

Various discussions with outstanding physics theoreticians of that
time such as Ya. B. Zeldovich, A. D. Sakharov, D. A.
Frank-Kamenetskii, Yu. B. Hariton, as well as members of the
experimental group of L. V. Altshuler were very essential for me.
I have already remarked that the comments of physicists were, as a
rule, very critical. They stimulated the detailed analysis of all
numerical effects and their reasons, and led to modifications of
the method. The specific problem for a steady flow around a body
was suggested to me by G. I. Petrov.

I wanted to demonstrate how the set of the principal scientific
issues emerged during that initial period of the development of
the computational fluid dynamics. Many of these issues are still
unresolved even today.

\section{References}
\qquad

1. G. B.~Alalikin, S. K.~Godunov, I. L.~Kireeva, and
L. A.~Pliner, {\em Solution of One-Dimensional Gas Dynamics Problems
Using Moving Grids\/}, (Nauka, Moscow, {\bf 1970}). [{\it In Russian\/}]

2. S.~K.~Godunov and A.~V.~Zabrodin, {\em On multi-dimensional
difference schemes of second order accuracy\/}, J. Comput. Math.
Math. Phys. 2(4), 706 (1962). [In Russian]

3. S.~K.~Godunov and K.~A.~Semendjaev, {\em Difference methods for
numerical solution of gas dynamics problems\/}, J. Comput. Math.
Math. Phys. 2(1), 3 (1962). [In Russian]

4. G.~P.~Prokopov and M.~V.~Stepanova, {\em The calculation of the
axisymmetric interaction of the interaction of a shock wave with a
blunt body at supersonic speeds\/}, preprint, IPM AN SSSR, No. 72,
1974. [In Russian]

5. S.~K.~Godunov, {\em Sur la construction des r\'{e}seaux dan les
domaines compliques d'une fa\c{}on automatique pour les \'equations
aux diff\'erences finies\/}, Act. Congress Int. Math. 3, 99 (1970).

6. S.K.~Godunov, A.V.~Zabrodin, and G.P.~Prokopov, {\em A difference
scheme for two-dimensional unsteady aerodynamics and an example of
calculations with a separated shock wave\/}, J. Comput. Math. Math.
Phys. 2(6), 1020 (1961). [In Russian]

7. S.~K.~Godunov and G.~P.~Prokopov, {\em On the solution of a discrete
Laplace's equation\/}, J. Comput. Math. Math. Phys. 9(2), 462 (1969).
[In Russian]

8. G.~P.~Prokopov, {\em On numerical implementation of the Vladimirov's
variational principle\/}, J. Comput. Math. Math. Phys. 8(1), 228
(1968). [In Russian]

9. S.~K.~Godunov, V.~V.~Ogneva, and G.~P.~Prokopov, {\em On the
convergence of the steepest descend method for the eigenvalue
problems\/}, in Partial Differential Equations, Proceedings of the
Symposium on 60th Anniversary of Academician S. L. Sobolev (Nauka,
Moscow, 1970). [In Russian]

10. S.~K.~Godunov and G.~P.~Prokopov, {\em Application of the method of
minimal iterations for the evaluation of the eigenvalues of elliptic
problems\/}, J. Comput. Math. Math. Phys. 10(5), 1180 (1970). [In
Russian]

11. S.~K.~Godunov, {\em Modern Aspects of Linear Algebra\/}, (Nauchnaja
Kniga, Novosibirsk, 1997) [In Russian]; English translation, Modern
Aspects of Linear Algebra, Translations of Mathematical Monographs
(Amer. Math. Soc., Pro\-vi\-dence, 1998), Vol. 175.

12. S. K. Godunov, {\em The problem of guaranteed precision of numerical
methods of linear algebra\/}, in Proc. Int. Congr. Math., Berkeley, CA,
1986, Vol. 2, p. 1353.

13. S.~K.~Godunov, A.~G.~Antonov, O.~P.~Kiriluk, and V.~I.~Kostin,
{\em Guaranteed Accuracy in Numerical Linear Algebra\/}, (Kluwer Academic,
Dordrecht, 1993); Rus\-sian edition published in 1992.

14. P. D. Lax, {\em Weak solutions of nonlinear hyperbolic equations and
their numerical computation\/}, Comm. Pure Appl. Math. 7(1), 159
(1954).

15. S.~K.~Godunov, {\em A difference scheme for numerical computation of
dis\-con\-tinuous solution of hydrodynamic equations\/}, Math. Sb. 47,
271 (1959) [In Russian]; translation, U.S. Joint Publ. Res. Service,
JPRS 7226 (1969).

16. S.~K.~Godunov, {\em Residues estimations for the approximate
solutions of the simplest equations of gasdynamics\/}, J. Comput. Math.
Math. Phys. 1(4), 622 (1961). [In Russian]

17. J.~Glimm, {\em Solutions in the large for nonlinear hyperbolic
systems of equa\-tions\/}, Comm. Pure Appl. Math. 18, 697 (1965).

18. B.~Van~Leer, {\em Towards the ultimate conservative difference
scheme. V. A second order sequel to Godunov's methods\/}, J. Comput.
Phys. 32, 101 (1979).

19. A. Harten, {\em A high resolution schemes for hyperbolic conservation
laws\/}, J. Comput. Phys. 49, 357 (1983).

20. A. Harten and P. D. Lax, {\em A random choice finite difference
scheme for hyperbolic conservation laws\/}, J. Numer. Anal. 18(2), 289
(1981).

21. A. Harten, {\em A high resolution schemes for hyperbolic conservation
laws\/}, J. Comput. Phys. 49, 357 (1983).

22. S. K. Godunov, {\em Lois de conservation et integrales d'\'energie
des \'equations hyperboliques\/}, in Nonlinear Hyperbolic Problems,
Proc. of Workshop, St.-Etienne, France, Jan. 13--19, 1988, Lecture
Notes in Mathematics (Springer-Verlag, New York, 1988), Vol. 1270,
p. 135.

23. S.~K.~Godunov, {\em Elements of Mechanics of Continuous Media\/}, (Nauka,
Mos\-cow, 1978). [In Russian]

24. K.~A.~Bagrinovskii and S.~K.~Godunov, {\em Difference schemes for
one-di\-men\-sional problems\/}, Dokl. Akad. Nauk SSSR 115(3), 431
(1957). [In Russian]

25. S.~K.~Godunov and G.~P.~Prokopov, {\em On the computations of
conformal mappings and grid generation\/}, J. Comput. Math. Math. Phys.
7(5), 1031 (1967). [In Russian]

26. S.~K.~Godunov and G.~P.~Prokopov, {\em On the utilization of moving
grids in gasdynamics computations\/}, J. Comput. Math. Math. Phys.
12(2), 429 (1972). [In Russian]

27. S.~K.~Godunov, V.~M.~Gordienko, and G.~A.~Chumakov, {\em Variational
prin\-ciple for 2-D regular quasi-isometric grid generation\/}, Int. J.
Comp. Fluid Dynamics 1995, V5, N1-2, pp. 99--118.

28. S.~K.~Godunov, A.~A.~Deribas, A.~V.~Zabrodin, and N.~S.~Kozin,
{\em Hydro\-dy\-namics effects in colliding solids\/}, J. Comput. Phys. 5,
517 (1970).

29. M. J. Ivanov and A. N. Kraiko, {\em Calculation of a transient
(sub-super) nozzle flows\/}, in Transactions of the Section for
Numerical Methods in Gasdynamics of the Second International Meeting
on the Gasdynamics of Explosion and Reacting Flows, Novosibirsk,
Moscow, 1969 (Computational Center of AN SSSR, 1971), p.3.[In
Russian]

30. T.~D.~Taylor and B.~S.~Mason, {\em Application of the unsteady
numerical method of Godunov to computation of supersonic flows past
bell shaped bodies\/}, J. Comput. Phys. 5(3), 443 (1970).

31. Numerical Solution of Multidimensional Gas Dynamics Problems,
edited by S. K. Godunov (Nauka, Moscow, 1976) [In Russian];
{\em R\'esolution num\'erique des probl\`emes multidimensionnels de la
dynamiqe des gaz\/}, edited by S. Godunov (Mir, Moscow, 1979). [In
French]

32. I.~M.~Gelfand, {\em Lectures on quasilinear equations\/}, Uspekhi Mat.
Nauk 14(2), 87 (1959). [In Russian]

33. S.~K.~Godunov, {\em Thermodynamics of gases and differential
equations\/}, Us\-pekhi Mat. Nauk 14(5), 97 (1959). [In Russian]

34. V.~F.~Diachenko, {\em On Cauchy problem for quasilinear equations\/},
Dokl. Akad. Nauk SSSR 136(1), 16 (1961). [In Russian]

35. S.~K.~Godunov, {\em On the concept of generalized solution\/}, Dokl.
Akad. Nauk SSSR 134(6), 1279 (1960). [In Russian]

36. S.~K.~Godunov, {\em On non unique "smearing" of discontinuities of
the solutions of quasi-linear systems\/}, Dokl. Akad. Nauk SSSR 136(2),
272 (1960). [In Russian]

37. K.~O.~Friedrichs and P.~D.~Lax, {\em Systems of conservation
equations with a convex extension\/}, Proc. Math. Acad. Sci. U.S.A.
68(8), 1696 (1971).

38. K. O. Friedrichs, {\em Conservation equations and the laws of motion
in classical physics\/}, Comm. Pure Appl. Math. 31, 123 (1978).

39. S. K. Godunov and E. I. Romensky, {\em Thermodynamical foundations
for special evolution differential equations of continuous media\/}, in
Trends in Appli\-cations of Mathematics to Mechanics (Longman,
Harlow/New York, 1995).

40. S. K. Godunov and E. I. Romenskii, {\em Thermodynamic, conservation
laws and symmetric forms of differential equations in mechanics of
continuous media\/}, in Computational Fluid Dynamics Review, 1995,
edited by M. Hafez and K. Oshima (Wiley, New York, 1995), p. 19.

41. K. S. Godunov and E. I. Romensky, {\em Symmetric forms of
thermodynamically compatible systems of conservation laws in
continuum mechanics\/}, in Numerical Methods in Engineering, Proc.
Second ECCOMAS Conf., Paris, France, 1996 (Wiley, Sons. New York,
1996).

42. S. K. Godunov, T. Y. Mikhailova, and E. I. Romenskii, {\em Systems of
ther\-mo\-dy\-na\-mically coordinated laws of conservation invariant
under rotations\/}, Siberian Math. J. 37(4), 690 (1996). [In Russian]

43. M. Sever, {\em Estimate of the time rate of entropy dissipation for
systems of conservation laws\/}, J. Differential Equations 130, 127.

\newpage

\begin{center} \textbf{\Large Appendix}  \end{center}

\textbf{ \large Experimental study of a disparity between order of
approximation and accuracy} \footnote{In this Appendix the results
obtained by V.V. Ostapenko (Lavrentev Institute of Hydrodynamics
SO RAS, Novosibirsk) are used.}\\[.4cm]

We present a method allowing an efficient estimation of  the order of 
the weak convergence, which is based on generalized solutions of
hyperbolic systems of conservation laws.  The idea of the method was
suggested by S. K. Godunov and V. S. Ryabenkii. The method is
based on the experimental estimation of the convergence rate of
the first integrals of numerical solutions which are calculated
over domains with singularities of the approximated exact
solution. The analysis of the accuracy  of the Harten's TVD-scheme
with the second order approximation on smooth solutions shows that
only the first order of convergence occurs in the problem of dam
breaking  with creation of a bore and a smooth rarefaction wave.
It results in the decreasing of the order of strong convergence in
smooth parts of exact solutions behind the wave front. The result
obtained is in contradiction with a wide-spread opinion according
to which the monotonic numerical schemes of higher order of
approximation on the smooth solutions conserve a higher order of
local convergence on smooth solutions of quasi-linear hyperbolic
conservation laws.

Let us consider the Cauchy problem for hyperbolic system of
quasi-linear con\-ser\-vation laws
$$
u_t + f(u)_x = 0\ ,\ \qquad  
u(0,x) = u_0(x)\ ,\ \qquad  
x \in \mathbb{R}\ , \eqno(1)
$$
where $u_0(x)$ is a $m$-dimensional piecewise continuous
vector-function and $f(u)$ a smooth flux function. We suppose that
the problem (1) has a unique stable ge\-ne\-ra\-lized solution
$u(t,x)$. The numerical solution is calculated with an explicit
two-layer in time and symmetric in space conservative scheme
$$
v_{j}^{n+1} = v_{j}^{n} 
            - \lambda_{n}\, ( \bar{f}_{j+1/2}^{n} 
                            - \bar{f}_{j-1/2}^{n} )\ ,\  \qquad 
  v_{j}^{0} = u_{0}(j\,h)\ ,\ \qquad j \in \mathbb{Z}\ ,\ \eqno(2)
$$
where 
$\bar{f}_{j+1/2}^{n} = \bar{f}(v_{j-k+1}^{n}, \dots
,v_{j+k}^{n})$,
$\bar{f}(u,\dots,u)=f(u)$, for all $ u \in \mathbb{R}^m$, 
$ v_{j}^{n} = v(t_n,j\,h)$, $t_0=0$, 
$t_n = \tau_0 + \tau_1 + \dots +\tau_{n-1}$, 
$\lambda_n = \tau_n / h$, $\tau_n $ is the time step at $n$-th time
layer $t_n$, $h$ is the  constant spatial step,
$\bar{f}$ is the  continuous function of numerical flux. The time
step $\tau_n$ is calculated from the Courant condition $\tau_n
\leq \tau^{\max}_n = z\, h /\max_{i,j} |a^i(v^n_j)|$, where $z=0.5$ is
a safety factor, $a^i(u)$, $i=1,\dots,m$ are the eigenvalues of Jacobi
matrix $f_u$ in (1). Let us fix a number $a$, $a \in\mathbb{R}$ and
introduce the integrals
$$
U^a(T,x) = \int_a^x u(T,y)\, \dd y\ ,\ \qquad 
V_h^a(T,x)= \int_a^x v_h(T,y)\, \dd y\ .
\eqno(3)
$$
The numerical solution $v_h(T,x)$ converges weakly to the exact
solution $u(T,x)$ with $r$-order within segment
$[a,x]\subset\mathbb{R}$ if
$$
V_h^a(T,x) - U^a(T,x) = C\, h^r + o(h^r), \eqno(4)
$$
where $C$ does not depend on $h$.

If $[a,x]$ does not contain singularities of the exact solution,
the order of weak convergence coincides with the order of a local
convergence on smooth solutions, i.e. equals to 2 in our case.
Otherwise, the order of weak convergence is less than 2, because
the TVD schemes do not have second order of weak approximation on
discontinuous functions.

Let us estimate the order of weak convergence $r$ in the case
where an exact discontinuous solution of the problem (1) is
unknown in advance. To calculate the order of weak convergence
(based on the Runge's rule) it is enough to have three numerical
results with rather small spatial steps $h_1=h$,$ h_2 = h/2$, $h_3=h/4$.
For each step the property (4) is satisfied, hence
$$
\delta V_i = |V^a_{h_i} - V^a_{h_{i+1}}| = |C|\, ( h^r_i - h^r_{i+1} )
+ o(h^r)\ ,\ \qquad i=1,2. \eqno(5)
$$
Therefore, we obtain with the accuracy $o(h^r)$
$$
\bfrac{\delta V_1}{\delta V_2}
=\bfrac{h_1^r - h_2^r}{h_2^r - h_3^r}
=\bfrac{1 - ( 1/2)^r }{ (1/2)^r - (1/4)^r }
= 2^r\ , 
$$
from which it follows that
$$
r = \log_2 \bfrac{\delta V_1}{\delta V_2}\ . \eqno(6)
$$

Below we present the results of a numerical calculation of errors
(5) and the order of accuracy (6) based on the TVD-scheme in the
problem of dam breaking  in a channel of non-constant depth.

As a test scheme one  chooses one of the TVD-schemes suggested by
Harten \footnote{J.Comp.Phys.1983, V.49, P.357-393.}. This scheme is
applied to the numerical solution of the Saint-Venant equations
(shallow water equations). They are equivalent to the equations of
the  polytropic gas with polytropic exponent $\gamma=2$ and have
the following form
$$
\begin{array}{lll}
  u_t + f(u)_x = 0\ , 
& u = \left( \begin{array}{c} 
              H \\ 
              q  
             \end{array}\right)\ , 
& f(u) = \left( \begin{array}{c} 
              q \\ 
              q^2 / H  + g\, H^2/2
                \end{array} \right)\ ,
\end{array}
$$
where $H$ and $q$ are the depth and the flux, respectively,  $g$
is the gravity (in calculations $g=10$).

In the Figures 1-4 for $T=0.2$ one can observe the results of the
calculation of the Cauchy problem  with discontinuous initial data
$$
H(0,x) = \left\{ \begin{array}{ll}  10\ ,              &  x \le 3.5\ , \\ 
                                    2 - th( x - 5 )\ , &  x > 3.5\ , 
                 \end{array} 
         \right. 
\eqno(7)
$$
$$
q(0,x) = 0\ , \qquad x \in \mathbb{R}\ . \eqno(8)
$$
In these calculations, $h_1 = h =0.1$, $h_2 = h/2 =0.05$, $h_3 = h/4 =0.025$.

The results of the calculations are presented in Figure 1: circles
correspond to the basic grid with step $h_1$, the solid line to 
the grid with step $h_3$; the dotted line corresponds to  the
initial position of the water level given by function (7).

Figures 2--3 are the  error, as defined by (5), and the order of
accuracy given by relation (6); $a=12$ in relation (3). From these
plots, one can see that, when $ x > 6$, i.e. when the segment of
integration $[a,x]$ lies entirely in the smooth part of the solution
just before the shock wave front, the order of weak convergence $r(x)$
equals 2.  When $ x < 5$ and when $[a,x]$ contains the front of a
discontinuous wave, then $r(x) \approx 1$, and the TVD-scheme (2) has
approximately only the first order of accuracy for the Cauchy problem
(7)--(8). 

In Figure 4, the function
$$
r_0(x)= \left\{ \begin{array}{ll} 
                 R(x)\ , & R(x) \le 3\ , \\
                 3\ ,    & R(x)  > 3\ ,
                \end{array} \right.
$$
is plotted where
$$
R(x)=\log \bfrac{\delta v_1(x)}{\delta v_2(x)}\ , \qquad 
\delta v_i(x) = |v_{h_i}(T,x) - v_{h_{i+1}}(T,x)|\ , \qquad i=1,2\ ,
$$
that presents the order of the local strong convergence of the
numerical solution. As one can see from Figure 4,  the calculation
domain is divided into three isolated parts by singularities of
the exact solution (inside  every part the order of local
convergence is changed in a random  way). Inside each part there
is a different order of the local convergence. The order of the
local convergence of the TVD-scheme on different smooth parts is
quite different, and, generally speaking, less than the order of
approximation on the  smooth solutions (see Fig.4).

\begin{figure}[th]
\begin{center}
\epsfig{file=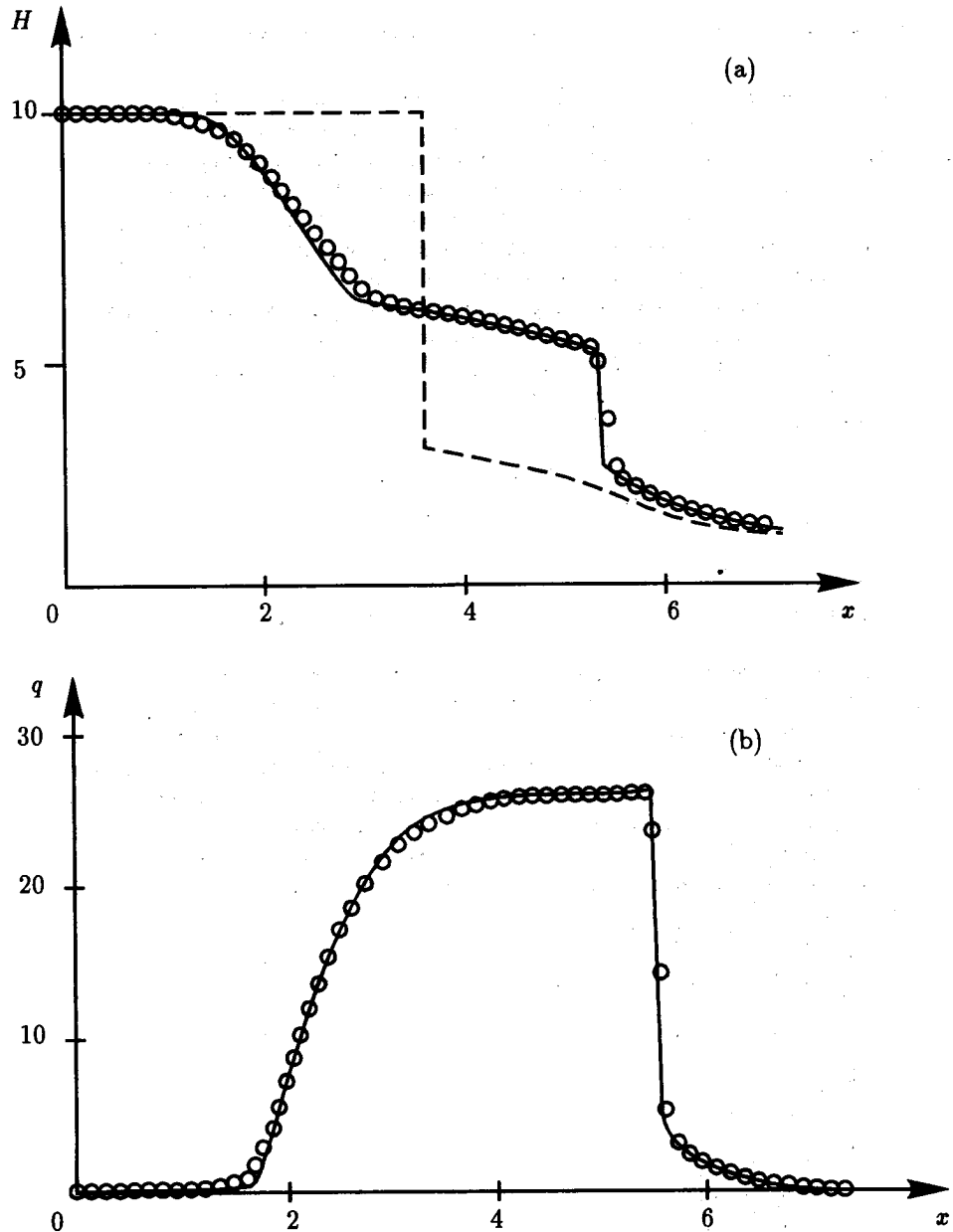}
\end{center}
\caption{Comparison of an exact and a numerical solution}
\label{1}
\end{figure}

\begin{figure}[th]
\begin{center}
\epsfig{file=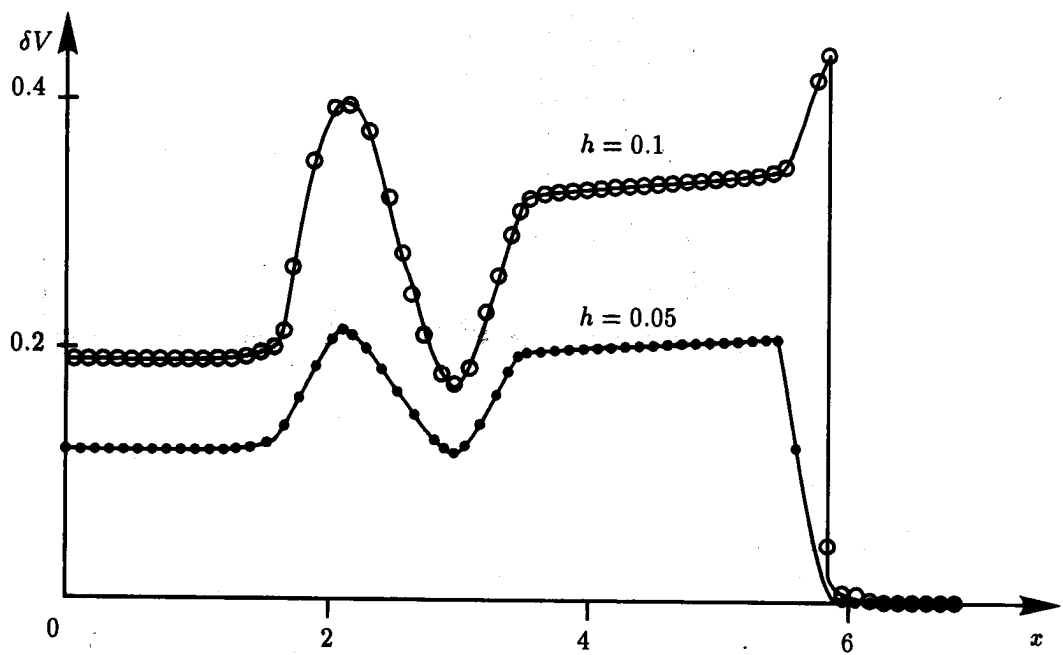}
\end{center}
\caption{Disbalances of the integrals of a numerical solution}
\label{2}
\end{figure}

\begin{figure}[th]
\begin{center}
\epsfig{file=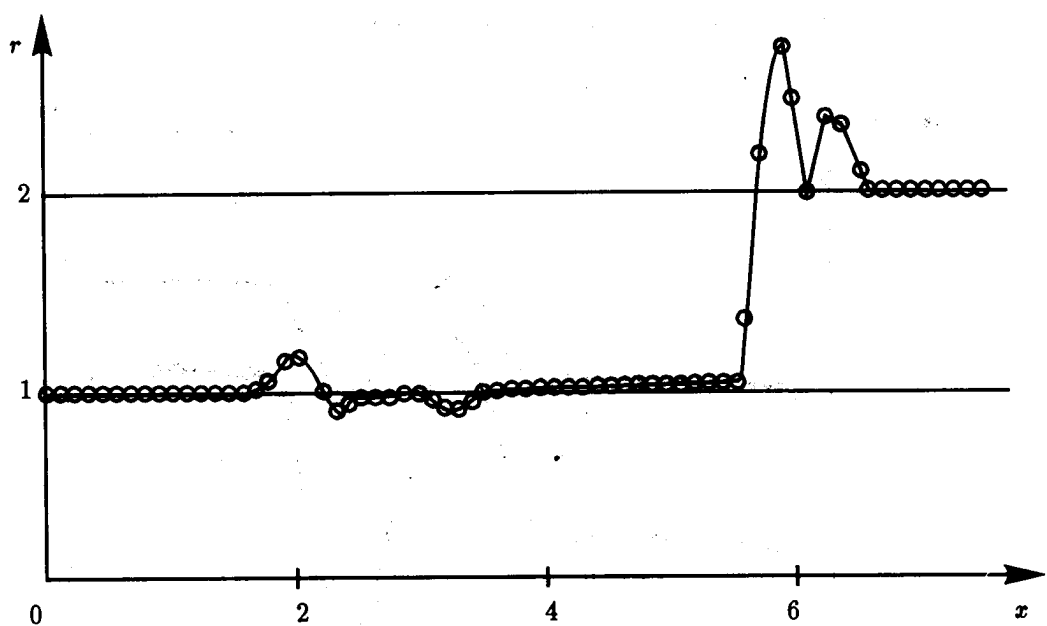}
\end{center}
\caption{The weak convergence order $r$} \label{3}
\end{figure}

\begin{figure}[th]
\begin{center}
\epsfig{file=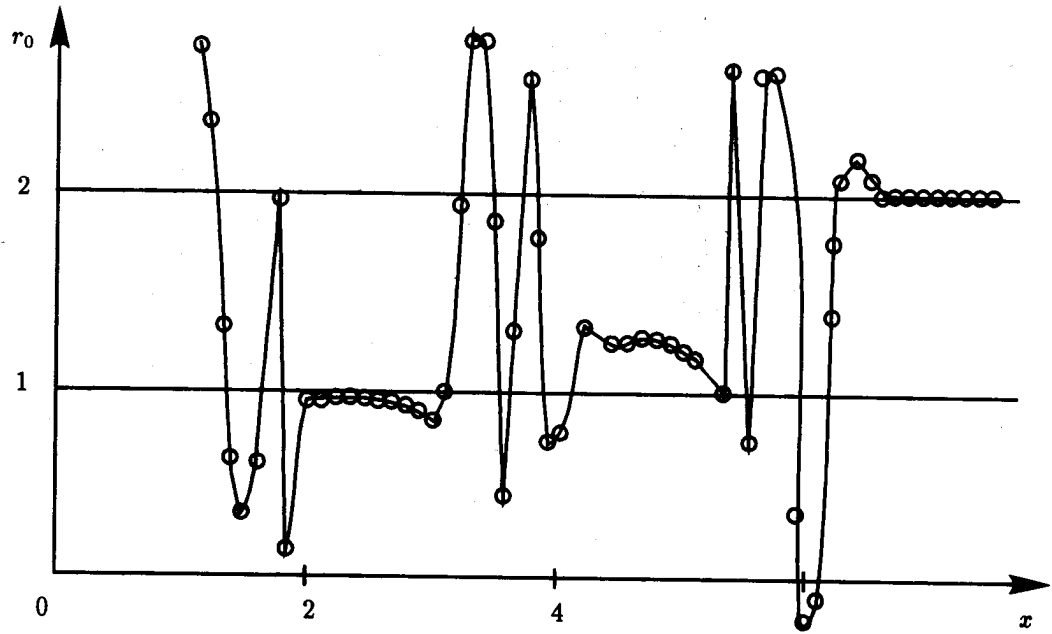}
\end{center}
\caption{The strong convergence order $r_0(x)$ } \label{4}
\end{figure}

\end{document}